\documentclass[epj]{svjour}
\usepackage{graphicx}
\usepackage{pslatex}
\usepackage{epstopdf}
\usepackage{color}
\usepackage{xspace}
\usepackage{arydshln}
\usepackage{soul} % use this (many fancier options)
\usepackage[normalem]{ulem}

\graphicspath{{figs/}}

\def\FastJet{{\sc FastJet}\xspace}
\def\mur{\ensuremath{\mu_r}\xspace}
\def\muf{\ensuremath{\mu_f}\xspace}
\def\mt{{m_t}}
\def\corr{\mbox{\scriptsize corr.}}

\def\mtPole{\ensuremath{M_t^{\mbox{\scriptsize pole}}}\xspace}
\def\mMS{\ensuremath{m_t(m_t)}\xspace}
\def\mMSmu{\ensuremath{m_t(\mu)}\xspace}

\def\GeV{\textnormal{GeV}}
\def\as{\ensuremath{\alpha_s}\xspace}
\def\ttbar{\ensuremath{t\bar t}\xspace}
\def\jet{\ensuremath{\mbox{jet}}}
\def\ttbaronejet{\ensuremath{t\bar t + 1\textnormal{\normalsize -jet}}\xspace}
\def\sigmattj{\ensuremath{\sigma_{t\bar t + \textnormal{\scriptsize
        1-jet}}}\xspace}
\def\sttj{\ensuremath{s_{t\bar t j}}\xspace}
\def\n3{\ensuremath{{\cal R}}\xspace}
\def\rhos{\ensuremath{\rho_s}\xspace}
\def\pt{\ensuremath{p_T}\xspace}
\def\MSbar{\ensuremath{\overline{\mbox{MS}}}\xspace}

\def\Fig#1{Fig.~\ref{#1}}
\def\Figs#1{Figs.~\ref{#1}}
\def\Eq#1{Eq.~(\ref{#1})}
\def\Ref#1{Ref.~\cite{#1}}
\def\Refs#1{Refs.~\cite{#1}}
\def\nn{\nonumber}

\def\makeheadbox{{%
    \hbox to0pt{\vbox{\baselineskip=10dd\hrule\hbox
        to\hsize{\vrule\kern3pt\vbox{\kern3pt
            \hbox{\sffamily HU-EP-17/05,  LAL 17-034, DESY 17-053}
            \kern3pt}\hfil\kern3pt\vrule}\hrule}%
      \hss}}}

\begin{document}
\titlerunning{Extracting the top-quark running mass using 
  $t\bar t + \mbox{1-jet}$ events.}

\title{Extracting the top-quark running mass using 
  $t\bar t + \mbox{1-jet}$ events produced at the Large
  Hadron Collider.}

\author{ J. Fuster\inst{1} \and A.
  Irles\inst{2,3} \and D. Melini \inst{1,4} \and P. Uwer\inst{5} \and M.
  Vos \inst{1} }

\authorrunning{J.~Fuster et al}
\institute{
  IFIC, Universitat de Val\`encia and CSIC, Catedr\'atico Jose
  Beltr\'an 2, 46980 Paterna, Spain 
  \and 
  Laboratoire de l'Acc\'el\'erateur Lin\'eaire, Centre Scientifique
  d'Orsay, Universit\'e de Paris-Sud XI, 
  CNRS/IN2P3, 91898 Orsay Cedex, France
  \and
  Deutsches Elektronen-Synchrotron (DESY), 
  Notkestra\ss{}e 85, 22607, Hamburg, Germany
  \and
  Universidad de Granada, Departamento de Fisica Te\'orica y del Cosmos,
  Campus Fuentenueva, Universidad de Granada, E-18071 Granada, Spain
  \and 
  Humboldt-Universit\"at zu Berlin, Newtonstrasse 15, 12489
  Berlin, Germany
}
\date{}

\abstract{We present the calculation of the next-to-leading order QCD
  corrections for top-quark pair production in association with an
  additional jet at hadron colliders, using the modified minimal
  subtraction scheme to renormalize the top-quark mass. The results
  are compared to measurements at the Large Hadron Collider run I. In
  particular, we determine the top-quark running mass from a fit of
  the theoretical results presented here to the LHC data.}


\maketitle

\section{Introduction}
\label{intro}
The top quark is the heaviest elementary particle discovered so
far. More than 25 years after its discovery at Fermilab 
\cite{Abe:1995hr,Abachi:1995iq} a
detailed understanding of why the top quark is so heavy is still
lacking. With a mass of 
\begin{equation}
  \label{massvalue}
  \mt = 173.34 \pm 0.27 \mbox{ (stat.)} \pm 0.71 \mbox{ (syst.)
  GeV~\cite{ATLAS:2014wva}}
\end{equation}
the top quark is roughly as heavy as a gold atom and more than 30 times
heavier than the next comparable heavy quark, the bottom quark. While the
top-quark mass may seem unnaturally large compared to the lighter
quark masses it appears rather natural given the size of the top-quark
Yukawa coupling which is approximately one.  Within the Standard Model
(SM) of particle physics, the top-quark mass is an important input
parameter influencing a variety of theoretical predictions:
\begin{enumerate}
\item In the SM the W-boson mass, the Higgs boson mass and the
  top-quark mass are related. A precise measurement of the
  three masses thus gives an important test of the SM.
\item The vacuum stability, analyzed through the behavior of the Higgs
  boson effective potential, is very sensitive to the top-quark mass.
  Given the current measurements, recent analysis show that the vacuum
  is most likely metastable  -  with a lifetime, however, larger than the
  age of the universe \cite{Degrassi:2012ry,Alekhin:2012py}.
\item Despite the large mass gap between the top-quark mass and the
  bottom-quark mass, the top quark highly influences the properties of
  B hadrons.
\end{enumerate}
While not being exhaustive the aforementioned examples nicely
illustrate the importance of precise determinations of the top-quark
mass. Tremendous efforts have been made in the past to improve
existing methods and to develop new approaches. For an overview we
refer to \Refs{Cortiana:2015rca,Vos:2016tof}.  Because of its short
lifetime the top quark is not observed as a free particle. As a
consequence, all top-quark mass determinations rely to some extent on
indirect determinations: the mass value is inferred by comparing the
measurements with theoretical predictions depending on the top-quark
mass.  Obviously, the accuracy of the procedure depends on the quality
of the measurements as well as on the uncertainties of the theoretical
predictions. Since QCD corrections can easily lead to corrections of
the order of ten per cent, at least next-to-leading order (NLO)
corrections should be taken into account.  Including radiative
corrections allows one also to give a unique interpretation of the
determined mass value within a specific renormalization scheme. In
particular, this gives the opportunity to consider other
renormalization schemes than the usually adopted pole mass scheme.  In
fact, the mass quoted in \Eq{massvalue} does not correspond to a well-defined
renormalization scheme. Very often this mass is identified as
the so-called Monte Carlo mass, since template fits and kinematical
reconstruction use Monte Carlo predictions to determine the mass
value. Although a rigorous proof is lacking it is often assumed that
the mass value is very close to the pole mass.  Using alternative
methods in which the renormalization scheme is better controlled thus
may provide valuable cross checks. In addition, different schemes may
show different behavior within perturbation theory. The freedom to
choose the scheme can thus be exploited to improve the mass
determination in specific cases. For example, it is well known that
the threshold behavior of top-quark pair production in electron-positron annihilation is badly described within perturbation theory
when the top-quark pole mass is used (see for example
\Ref{Hoang:2000yr} and the references therein).  It is also well known
that the pole mass concept suffers from the so-called renormalon
ambiguity reflecting the fact that the pole mass is strictly speaking
not well defined in QCD because of confinement
\cite{Bigi:1994em,Beneke:1994sw}.  While the related uncertainty of
order $\Lambda_{\rm QCD}$\footnote{Recent work indicates that the
  uncertainty is below 100 MeV \cite{Beneke:2016cbu}.} might still be
negligible in view of the precision reached in current measurements,
it is nevertheless highly interesting and well motivated to
investigate also top-quark mass measurements using alternative mass
definitions. 
This has been done for the first time in
\Ref{Langenfeld:2009wd} where the determination of the top-quark mass
in the modified minimal subtraction/running mass scheme (\MSbar) has
been studied. The analysis is based on cross section measurements at
the Tevatron.  Using in the theoretical predictions the \MSbar mass
instead of the pole mass, the comparison with the experimentally
measured cross sections gives direct access to the running mass. The
results obtained in \Ref{Langenfeld:2009wd} and subsequent
determinations are consistent with the pole mass measurements. Similar
studies have been performed at the LHC.  Very recently, the analysis
has been extended to single top-quark production
\cite{Alekhin:2016jjz}.  

In \Ref{Alioli:2013mxa} an alternative
observable to determine the top-quark mass has been proposed.  In this
case, top-quark pairs in association with an additional jet are
considered instead of the inclusive cross section for top-quark pair
production. Since the emission of an additional jet also depends on
the top-quark mass, this process has a high sensitivity to the mass
parameter.  Indeed, it has been shown in \Ref{Alioli:2013mxa} that the
invariant mass distribution of the $t\bar t-\jet$ system
significantly enhances the mass effects. In
\Refs{Aad:2015waa,CMS:2016khu} the method has been employed to
determine the top-quark pole mass using LHC run I data.  The aim of
this article is to extend the calculation of
\Ref{Alioli:2013mxa} to the \MSbar scheme and apply the method to
extract the running mass \mMS using published results from the ATLAS
experiment \cite{Aad:2015waa}.  This may serve as a proof of concept
as well as a consistency check of the existing measurement.  The
article is organized as follows. In Sect. \ref{sec:theory}
theoretical results using the running top-quark mass are presented.
Using these results we determine in Sect.
\ref{sec:massdetermination} the running top-quark mass from a
comparison with published experimental results \cite{Aad:2015waa}. We
finally close with a short conclusion in Sect. \ref{sec:conclusion}.

\section{Theoretical predictions using the \MSbar mass}
\label{sec:theory}
We study the process
\begin{equation}
  pp \to t\bar t + \mbox{1-jet} + X.
\end{equation}
Note that a minimal \pt of the additional jet is required to separate
the process from inclusive top-quark pair production and render the
cross section infrared finite.  In the following, we adopt the 50 GeV
cut also applied in the experimental analysis \cite{Aad:2015waa}.  The
NLO corrections for top-quark pair production in association with an
additional jet have been calculated first in
\Refs{Dittmaier:2007wz,Dittmaier:2008uj} and later in
\Refs{Melnikov:2009dn,Melnikov:2010iu}. In
\Refs{Dittmaier:2007wz,Dittmaier:2008uj} as well as in
\Refs{Melnikov:2009dn,Melnikov:2010iu} the top-quark mass is
renormalized in the pole mass scheme. In the pole mass scheme the
top-quark mass $\mtPole$ is defined as the location of the pole of the
perturbatively calculable top-quark propagator. In
\Refs{Dittmaier:2007wz,Dittmaier:2008uj} it has been shown that the
QCD corrections, although not negligible, are moderate in size. In
particular, the scale dependence is stabilized in NLO and shows a
plateau at $\mur=\mtPole$ where $\mur$ denotes the renormalization
scale which is set equal to the factorization scale $\muf$. In
\Ref{Alioli:2013mxa} these findings have been extended by the
observation that also different approximations in the theoretical
description (fixed order, $t\bar t+X$ + parton shower,
$\ttbaronejet+X$ + parton shower), using again the top-quark pole mass,
show remarkably consistent results.  Given the stability with respect
to radiative corrections, it has been argued in \Ref{Alioli:2013mxa}
that top-quark pair production in association with an additional jet
may provide a promising alternative to measure the top-quark mass. As
an observable the quantity
\begin{equation}
  \label{eq:n3Definition}
  \n3(\mtPole,\rhos)=
  \frac{1}{\sigmattj} 
  \frac{d\sigmattj}{d\rhos}(\mtPole,\rhos)
\end{equation}
has been proposed.
Here $\sigmattj$ is the total cross section for $pp\to \ttbaronejet$
(including the \pt cut) and $\rhos$ is a dimensionless variable
defined as
\begin{equation}
  \rhos=\frac{2 m_0} {\sqrt{\sttj}},
\end{equation}
where $m_0$ is an `arbitrary' mass scale of the order of the top-quark
mass. In the following $m_0=170$~GeV is used.  The distribution
$\n3(\mtPole,\rhos)$ has been successfully used by the LHC experiments
ATLAS and CMS to determine the top-quark mass in the pole mass scheme
\cite{Aad:2015waa,CMS:2016khu}. Since the aim of this article is to
repeat the analysis using, however, the mass parameter renormalized in
the \MSbar scheme, theoretical predictions within this scheme are
required.  Owing to the Lehmann-Symanzik-Zimmermann formalism an ab
initio calculation using the \MSbar scheme is non-trivial. To obtain
results in the \MSbar scheme we thus follow the method developed in
\Refs{Langenfeld:2009wd}.  To illustrate the main idea and for convenience,
we briefly outline the approach for the cross section. The
same technique is also applicable to each individual bin of a
differential distribution.  The relation between the \MSbar
mass $m_t(\mu)$ and the pole mass $\mtPole$ is given by
\begin{equation}
  \label{eq:M1}
  \mtPole =  m_t(\mu)\left( 1 + \hat a(\mu) 
    {4\over 3}\left[1-{3\over 4}\ln\left({m_t^2\over
          \mu^2}\right)\right] 
  \right)  
  +O(\hat a^2)
\end{equation}
with 
\begin{equation}
  \hat a(\mu) =  {\as^{(6)}(\mu) \over \pi}
\end{equation}
where $\as^{(6)}(\mu)$ is the QCD coupling constant of the strong
interaction in the six flavor theory. In
\Ref{Dittmaier:2007wz,Dittmaier:2008uj} the closed top-quark loop has
been subtracted at finite momentum. At the order we are working here
the coupling constant defined in this way corresponds to the coupling
constant in the five flavor theory. Since the difference between
$\as^{(5)}(\mu)$ and $\as^{(6)}(\mu)$ is again a higher-order effect
we can replace in \Eq{eq:M1} $\hat a(\mu)$ by
\begin{equation}
  a(\mu) =  {\as^{(5)}(\mu) \over \pi}
\end{equation}
to obtain 
\begin{equation}
\mtPole =  \mMS\left( 1 + a(\mu) {4\over 3}\right) + O(a^2).
\end{equation}
Note that we have replaced $a(\mt)$ by $a(\mu)$, which is again a
higher-order effect.
In the pole mass scheme the expansion of the cross section $\sigma$
reads
\begin{equation}
  \sigma = a(\mu)^3 \sigma^{(0)}(\mtPole) 
  + a(\mu)^4 \sigma^{(1)}(\mtPole) + \ldots,
\end{equation}
where $\sigma^{(0)}$, $\sigma^{(1)}$ denote the expansion coefficients.
For simplicity, we have suppressed all further dependence of $\sigma^{(0)}$
and $\sigma^{(1)}$.
Using the relation between the pole mass and the \MSbar mass given above we
get
\begin{eqnarray}
  \sigma &=& a(\mu)^3 \sigma^{(0)}\left(\mMS(1+ {4\over 3} a(\mu) + \ldots
    )\right)\\ \nonumber
  &+& a(\mu)^4 \sigma^{(1)}\left(\mMS(1+ {4\over 3} a(\mu)  + \ldots
    )\right)
  + \ldots
\end{eqnarray}
Working in fixed-order perturbation theory we have to expand 
in the coupling constant:
\begin{eqnarray}
  \label{eq:SchemeConversion}
  \sigma &=& 
  a(\mu)^3  \sigma^{(0)}(\mMS)  + a(\mu)^4 \bigg[ \sigma^{(1)}(\mMS) \nn\\
  &+& {4\over 3}\mMS\left.{d\sigma^{(0)}(\mtPole)\over
        d\mtPole}\right|_{\mtPole=\mMS} 
  \bigg] + O(a^5).
\end{eqnarray}
The two contributions $ \sigma^{(0)}(\mMS)$, $\sigma^{(1)}(\mMS) $ are
trivial to compute: one just evaluates the corresponding cross sections
with the pole mass set to the numerical value of $\mMS$. The third term
is a bit more cumbersome. Instead of evaluating the derivative
analytically, we decided to compute it numerically. To do so we computed
$\sigma^{(0)}(\mMS\pm \Delta)$ and  $\sigma^{(0)}(\mMS\pm 2\Delta)$
and used
\begin{equation}
  \label{eq:TwoPoint}
  {df(x)\over dx} = {f(x+\Delta)-f(x-\Delta)\over 2\Delta} + O(\Delta^2)
\end{equation}
as well as 
\begin{eqnarray}
  {df(x)\over dx} &=& \frac{1}{12 \Delta}\bigg(
  f(x-2\Delta)- 8\*f(x-\Delta)\\\nn
  &+&8\*f(x+\Delta) - f(x+2\Delta)\bigg)
  + O(\Delta^4).
\end{eqnarray}
As a further consistency check, we used different
values for $\Delta$: 1, 0.5 and 0.25~$\GeV$. Using
different discretizations and different approximations lead to
differences for the cross section at the per mille level.
\begin{figure}
  \includegraphics[width=\columnwidth]{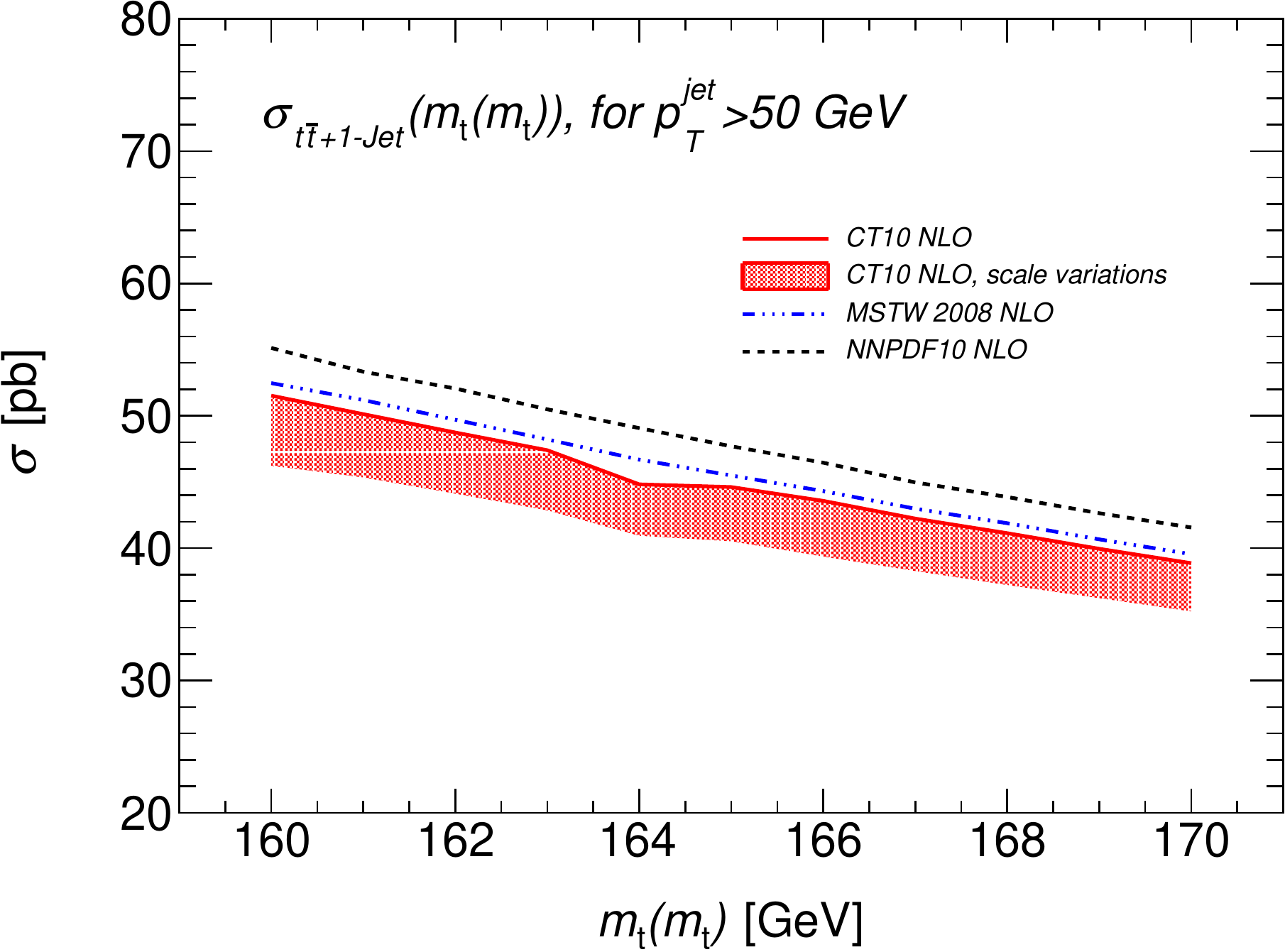}
  \caption{The $\ttbaronejet + X$ cross section at
    NLO QCD for
    proton-proton collisions at 7 TeV as a function of the running mass \mMS.
    The lines show the 
    result of the calculation for $\mu=\mMS$ using different PDF sets:
    CT10 NLO (red solid line), 
    MSTW2008nlo90cl (blue dashed-dotted line), 
    NNPDF2.3 NLO PDF (black dashed line).
    The red shaded area shows the impact of the 
    scale variation for the CT10 NLO PDF set, where $\mu=\mur=\muf$
    has been varied in the range
    $\mMS/2\le\mu\le2\mMS$.
    \label{fig:inclusive_cross}}
\end{figure}

When applying this method to the $\ttbaronejet + X$ cross section at
7~TeV as function of the \MSbar mass \mMS we obtain the results shown
in \Fig{fig:inclusive_cross}.  As in \Refs{Alioli:2013mxa,Aad:2015waa}
the jets are defined using the anti-kt algorithm
\cite{Cacciari:2008gp} as implemented in the \FastJet package
\cite{Cacciari:2011ma} with R=0.4 and a recombination according to the
$E-$scheme. Furthermore, as stated above the additional jet is
required to have $\pt>50$~GeV and in accordance with \Ref{Aad:2015waa}
$|\eta|<2.5$ where $\eta$ defines the pseudo-rapidity. As far as the
coordinate system is concerned, we follow the LHC convention which
defines the $z$-direction along the beam axis and the origin as the
nominal interaction point. For more details we refer to
\Ref{Aad:2015waa} (footnote below Eq.~(2)).  Having applied the same
cuts as in \Ref{Aad:2015waa} the results for \n3 presented here may be
directly compared with the unfolded results as presented in
\Ref{Aad:2015waa}.  

To study the PDF dependence of the theoretical predictions, three
different PDF sets are employed: the CT10 NLO PDF
set~\cite{Nadolsky:2008zw}, which is the nominal PDF set used for the
top-quark pole mass extraction in \Ref{Aad:2015waa}, the
MSTW2008nlo90cl PDF set~\cite{Martin:2009iq}, and the NNPDF2.3 NLO PDF
set~\cite{Ball:2012cx}. The MSTW2008 NLO and the NNPDF2.3 NLO sets are
chosen to follow as closely as possible the analysis performed in
\Ref{Aad:2015waa} where the two sets have been used to estimate the
PDF effects. CT10 NLO and MSTW2008 give rather consistent results,
deviating in most cases less than two per cent from each other.  The
NNPDF2.3 NLO PDF leads to larger results which differ by about four per
cent from MSTW2008. Given the progress concerning the PDF
determinations recent PDF sets should show for even smaller differences.

To estimate the effect of higher-order corrections we calculate for
the CT10 NLO PDF set the impact of the scale dependence varying the
scale as usual by a factor 2 up and down around the central scale
set to $\mu=\mMS$.  Similar to what has been observed in
\Refs{Dittmaier:2007wz,Dittmaier:2008uj} when using the pole mass, the
predictions for the cross section for $\mu=2\mMS$ and $\mu=0.5\mMS$
are both smaller than the result for $\mu=\mMS$, showing a plateau
around $\mu=\mMS$.  The band in~\Fig{fig:inclusive_cross} shows the
largest variation with respect to the nominal value for $\mu=\mMS$.
With respect to the central value of the band, the scale uncertainty
amounts to an effect of about $\pm 5\%$ - slightly smaller than
the scale uncertainty observed using the pole mass definition. This is
very similar to what is observed for the inclusive top-quark pair
production cross
section. Note that this is merely a kinematic effect as in the \MSbar
scheme the leading-order results increase because of the numerically
smaller mass value.  The NLO corrections decrease accordingly and thus
lead to a smaller scale dependence.

\begin{figure}
  \includegraphics[width=\columnwidth]{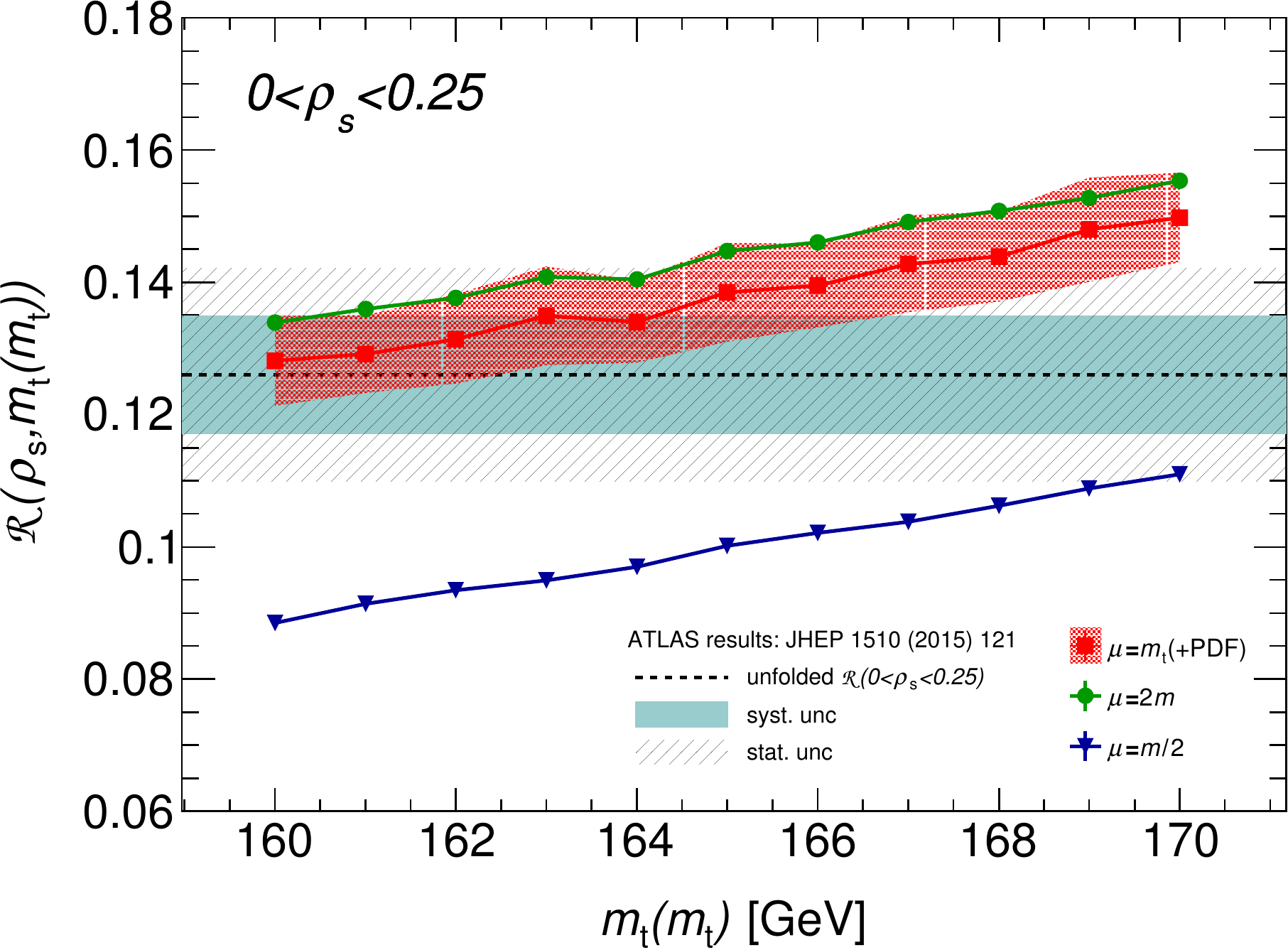}
  \includegraphics[width=\columnwidth]{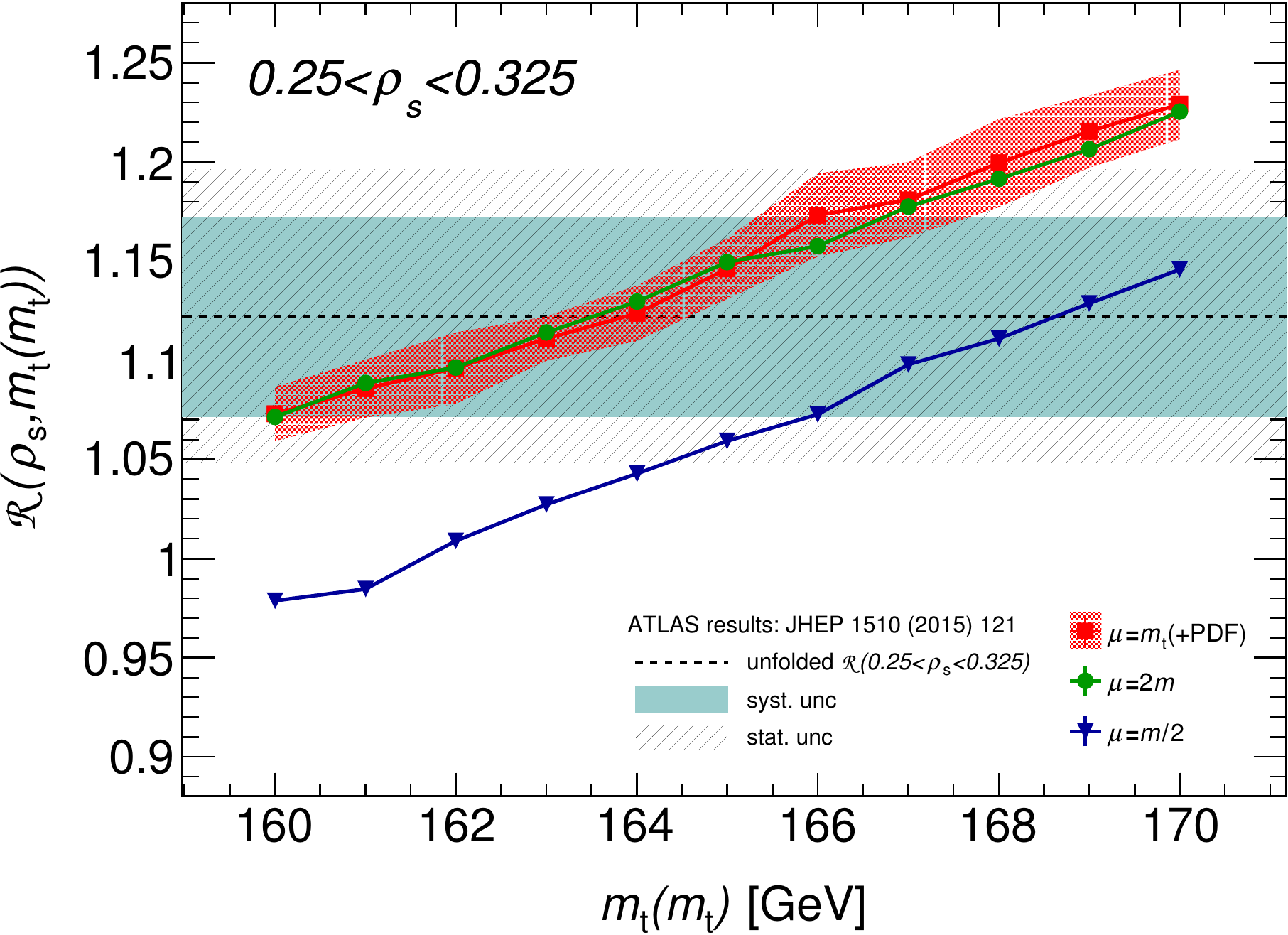}
  \includegraphics[width=\columnwidth]{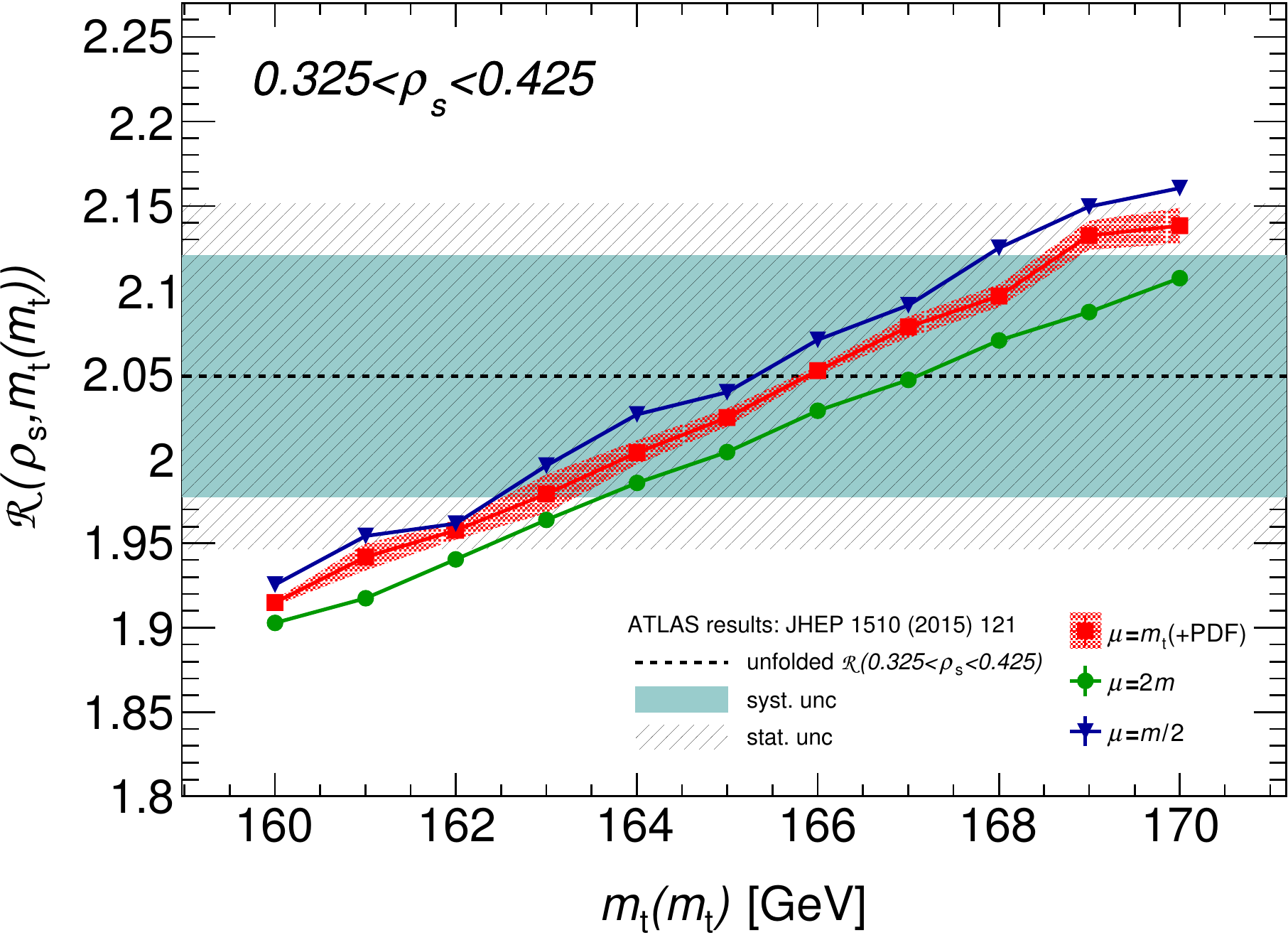}
  \caption{\n3 as defined in \Eq{eq:n3Definition} as a 
    function of \mMS for the $\rhos<0.425$ intervals using the CT10
    PDF set.  The red squares show the results for the central scale
    $\mu=\mur=\muf=\mMS$. The red band illustrates the impact of using
    the MSTW2008 or the NNPDF2.3 PDF set. The blue triangles (green
    dots) show the results for $\mu=\mMS/2$ ($\mu=2\mMS$). For
    comparison we also illustrated as dashed line the measured result
    as reported in \Ref{Aad:2015waa}. The bands show the systematic and
    statistical uncertainty.
    \label{fig:bin1}}
\end{figure}

\begin{figure}
  \includegraphics[width=\columnwidth]{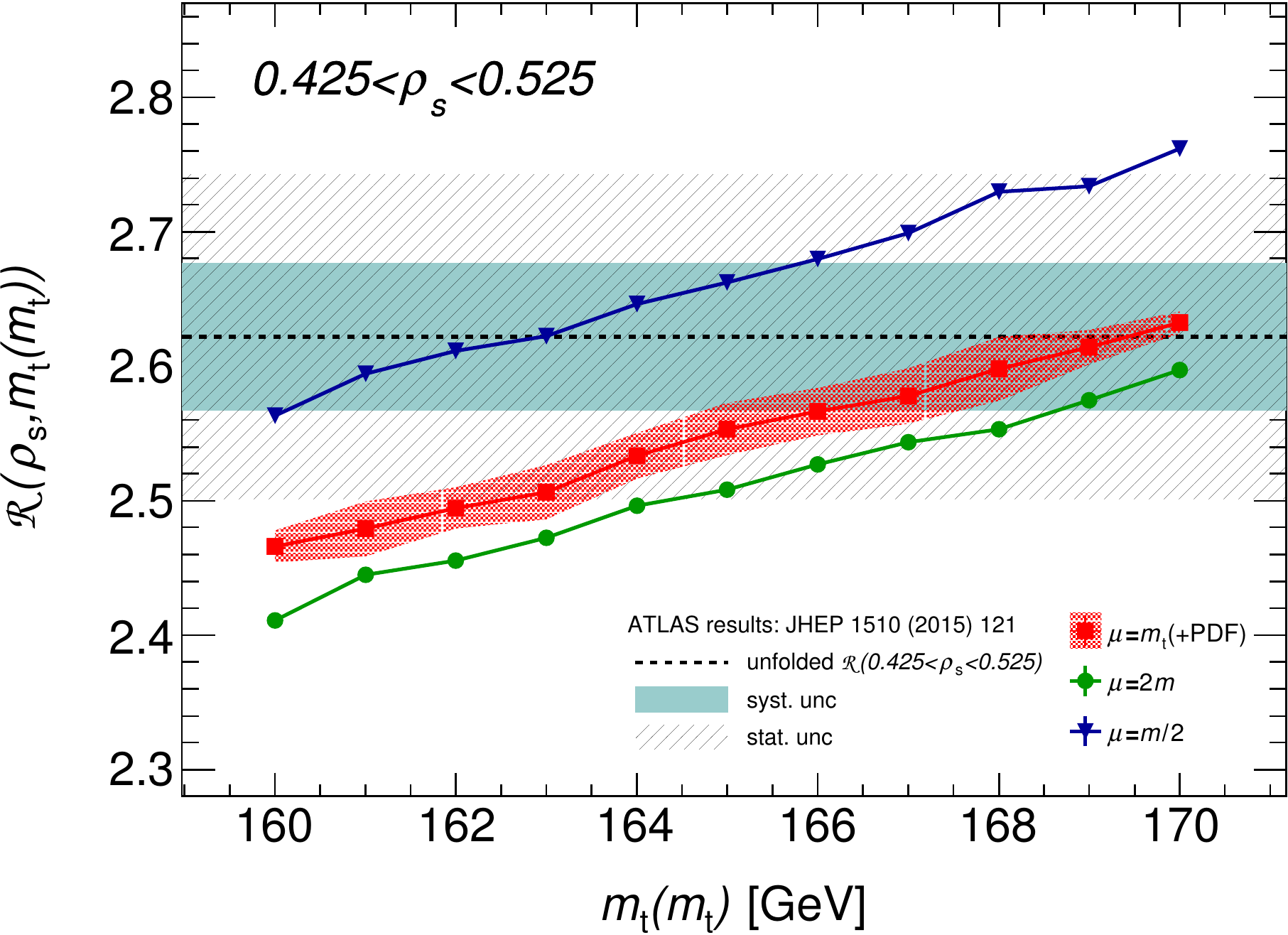}
  \includegraphics[width=\columnwidth]{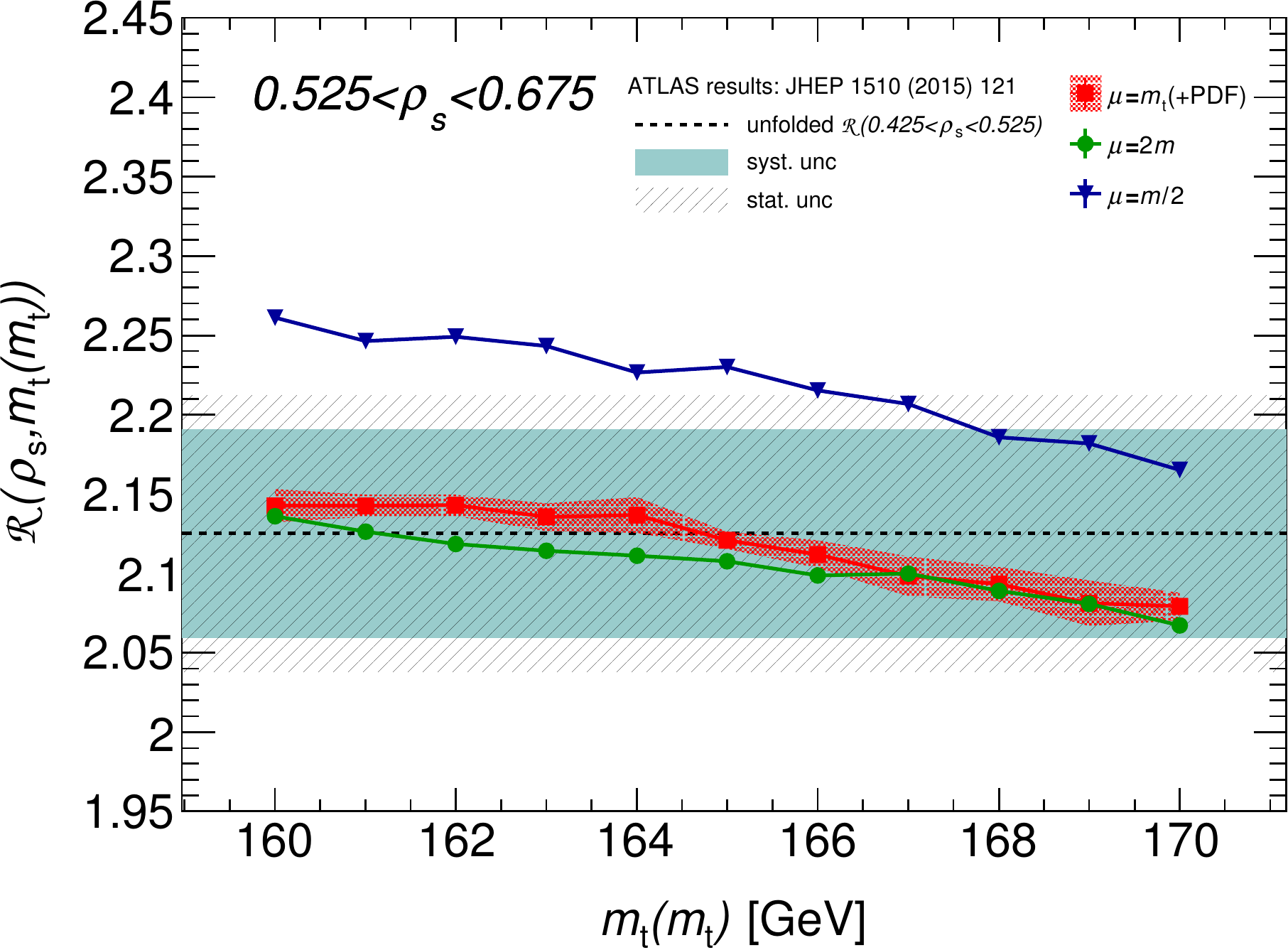}
  \includegraphics[width=\columnwidth]{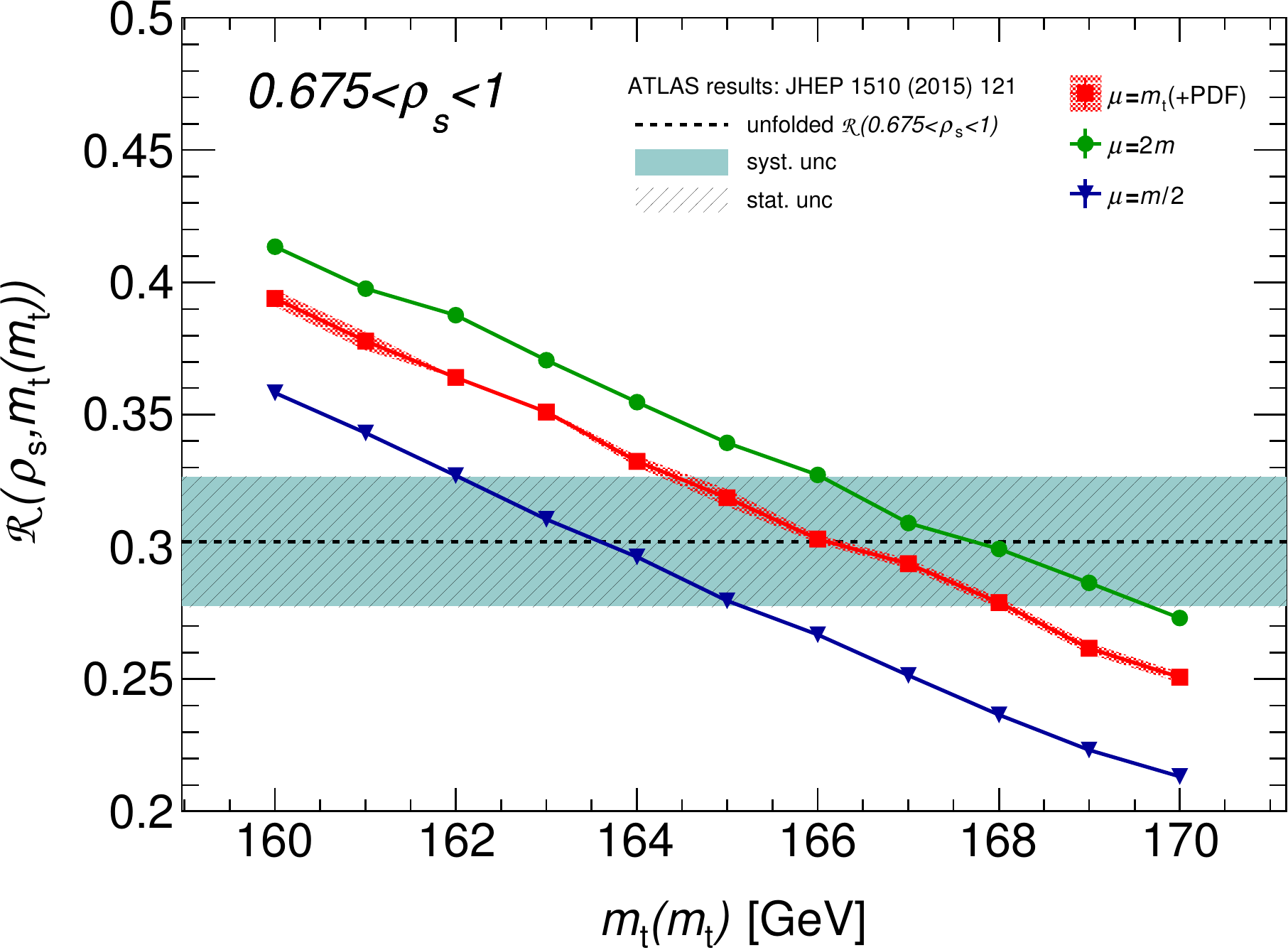}
  \caption{Same as \Fig{fig:bin1} but for the  $\rhos>0.425$ 
    intervals. \label{fig:bin4}}
\end{figure}

To calculate the $\cal R$ distribution defined in \Eq{eq:n3Definition}
using the \MSbar mass, we apply the same strategy as for the cross
section employing, however only the two-point formulas of
Eq.~(\ref{eq:TwoPoint}) together with a step size of $\Delta =0.5~\GeV$.
For one mass value we cross checked the procedure using the
five point formulas with a step size of 0.25~GeV. We found consistent
results at the per mille level.  The results for $\cal R$ for the
different bins in $\rhos$ are shown in \Figs{fig:bin1} and
\ref{fig:bin4}. The definition of the bin boundaries follows the setup
used in \Refs{Aad:2015waa}.  Again the calculation has been done for
three different scales $\mu=\mMS/2,\mMS,2\mMS$. The predictions for
${\cal R}$ for different values of $\mMS$ are given in
Table~\ref{tab:n3_scale}. The upper and lower subscripts denote the
shift with respect to the central scale $\mu=\mMS$ (the upper value
gives the shift for $\mu=2\mMS$, the lower value gives the shift for
$\mu=\mMS/2$). As for the $\ttbaronejet$ cross section we employ the
CT10 PDF set.  Results for the two other PDF sets used in this
analysis are given in the appendix in Table~\ref{tab:n3_pdf_nnpdf} for
the NNPDF PDF set and Table~\ref{tab:n3_pdf_mstw} for the MSTW PDF
set. In \Figs{fig:bin1} and \ref{fig:bin4} the red squares show the
results for the central scale, while the blue triangles (green
circles) show the results for $\mu=\mMS/2$ ($\mu=2\mMS$).  To
illustrate the PDF dependence, the dotted band shows the maximal shift
of the predictions with respect to the CT10 NLO PDF set using the
MSTW2008nlo90cl and NNPDF2.3 NLO PDF sets.  In addition to the
theoretical predictions, we show as horizontal bands for each bin in
$\rhos$ the experimental result for \n3 - unfolded to the parton
level - as published in Table 4 of \Ref{Aad:2015waa}. The dashed black
line illustrates the central values, while the hashed band shows the
statistical uncertainty and the solid band shows the systematic
uncertainty. The systematic uncertainties include theoretical and
experimental uncertainties. We stress that the results for \n3 given
in Table 4 of \Ref{Aad:2015waa} have been unfolded to the parton-level,
detector effects and hadronization effects are no longer accounted.

It is thus possible to directly compare the
measurements with the theoretical predictions presented here. In
particular, the theoretical results for \n3 as a function of the
top-quark \MSbar mass, can be fitted to the
unfolded distribution, leading to a determination of the top-quark
mass in the \MSbar scheme.
%%
%% Table with theoretical R-values
\begin{table}[t]
  \begin{center}\renewcommand{\arraystretch}{1.6}
    \begin{tabular}{l|l|l|l} 
      \hline
      & \multicolumn{3}{c}{$\n3(\mMS)$ }\\
      \hline
      Bin, range in $\rhos$  & $160$ GeV & $161$ GeV &  $162$ GeV   \\
      \hline                                                            
      1, 0-0.25   & $0.128^{+0.006}_{-0.040}$   &    
      $0.129^{+0.007}_{-0.038}$     &    $0.131^{+0.006}_{-0.038}$ \\
      2, 0.25-0.325  & $1.073^{-0.001}_{-0.094}$  &    
      $1.086^{+0.003}_{-0.101}$     &    $1.096^{+0.000}_{-0.087}$ \\
      3, 0.325-0.425 & $1.915^{-0.012}_{+0.011}$   &    
      $1.942^{-0.024}_{+0.012}$     &    $1.957^{-0.017}_{+0.004}$ \\
      4, 0.425-0.525 & $2.466^{-0.055}_{+0.097}$   &    
      $2.479^{-0.034}_{+0.115}$     &    $2.495^{-0.039}_{+0.117}$ \\
      5, 0.525-0.675  & $2.142^{-0.006}_{+0.119}$   &    
      $2.143^{-0.016}_{+0.104}$     &    $2.143^{-0.024}_{+0.106}$ \\
      6, 0.675-1.0    & $0.394^{+0.020}_{-0.036}$   &    
      $0.378^{+0.020}_{-0.035}$     &    $0.364^{+0.024}_{-0.037}$ \\
      \hline
    \end{tabular}\vspace{-0.5cm}
    \begin{tabular}{c|l|l|l|l} 
      \multicolumn{4}{c}{}\\
      \multicolumn{4}{c}{}\\
    \hline
    bin&    $163$ GeV &  $164$ GeV &  $165$ GeV &  $166$ GeV \\
    \hline                                                            
    1&$0.135^{+0.006}_{-0.040}$     &    $0.134^{+0.006}_{-0.037}$     
    &    $0.139^{+0.006}_{-0.038}$     &    $0.140^{+0.007}_{-0.037}$ \\
    2&$1.111^{+0.003}_{-0.084}$     &    $1.124^{+0.006}_{-0.081}$     
    &    $1.146^{+0.003}_{-0.087}$     &    $1.173^{-0.016}_{-0.101}$ \\
    3&$1.979^{-0.015}_{+0.017}$     &    $2.004^{-0.018}_{+0.023}$     
    &    $2.024^{-0.020}_{+0.015}$     &    $2.052^{-0.024}_{+0.018}$ \\
    4&$2.506^{-0.034}_{+0.116}$     &    $2.533^{-0.037}_{+0.113}$     
    &    $2.553^{-0.045}_{+0.109}$     &    $2.566^{-0.039}_{+0.114}$ \\
    5&$2.135^{-0.021}_{+0.108}$     &    $2.137^{-0.026}_{+0.090}$     
    &    $2.121^{-0.013}_{+0.109}$     &    $2.112^{-0.013}_{+0.103}$ \\
    6&$0.351^{+0.020}_{-0.040}$     &    $0.332^{+0.022}_{-0.036}$     
    &    $0.319^{+0.021}_{-0.039}$     &    $0.303^{+0.024}_{-0.036}$ \\
    \hline
  \end{tabular}\vspace{-0.5cm}
  \begin{tabular}{c|l|l|l|l} 
    \multicolumn{4}{c}{}\\
    \multicolumn{4}{c}{}\\
    \hline
    bin&$167$ GeV &  $168$ GeV &  $169$ GeV &  $170$ GeV  \\
    \hline                                                            
    1&$0.143^{+0.006}_{-0.039}$     &    $0.144^{+0.007}_{-0.038}$     
    &    $0.148^{+0.005}_{-0.039}$     &    $0.150^{+0.006}_{-0.039}$ \\
    2&$1.181^{-0.003}_{-0.083}$    &    $1.199^{-0.008}_{-0.089}$    
    &    $1.215^{-0.009}_{-0.087}$    &    $1.229^{-0.004}_{-0.083}$ \\
    3&$2.078^{-0.031}_{+0.013}$     &    $2.097^{-0.027}_{+0.028}$     
    &    $2.132^{-0.045}_{+0.017}$     &    $2.138^{-0.031}_{+0.022}$ \\
    4&$2.578^{-0.034}_{+0.121}$     &    $2.598^{-0.045}_{+0.132}$     
    &    $2.614^{-0.040}_{+0.120}$     &    $2.632^{-0.035}_{+0.130}$ \\
    5&$2.098^{+0.002}_{+0.109}$      &    $2.093^{-0.004}_{+0.092}$     
    &    $2.081^{-0.000}_{+0.101}$     &    $2.079^{-0.012}_{+0.086}$ \\
    6&$0.294^{+0.015}_{-0.042}$     &    $0.279^{+0.020}_{-0.042}$     
    &    $0.262^{+0.025}_{-0.038}$     &    $0.251^{+0.022}_{-0.037}$  \\
    \hline
  \end{tabular}
\end{center}
\caption{The $\n3(\mMS)$-distribution calculated using 
  \ttbaronejet samples at NLO accuracy for different $\mMS$ values
  for the CT10 PDF set. 
  The predictions for $\mu=\mt$ are quoted as central values. The
  shifts correspond to the difference ($\n3(\mt,\mu=x)-\n3(\mt,\mu=\mt)$) 
  where the results for $x=2\mt$ and $x=\mt/2$ 
  are quoted as upper-scripts and lower-scripts correspondingly.
  \label{tab:n3_scale}}
\end{table}
Since the analysis presented in this paper relies on the usage of
unfolded data we may add some remarks concerning the unfolding. 
Although the full reconstruction of the top quarks is not required we
stress that the reconstruction of the momenta of the final state
objects which is required to determine the invariant mass squared $\sttj$,
is highly non-trivial, since issues like missing energy, unobserved
neutrinos, and combinatorial background have to be
taken into account. In particular, the aforementioned issues will also
lead to an additional uncertainty of the unfolding procedure.
%% While the reconstruction of the top-quarks is not required we stress that
%% the unfolding procedure includes the reconstruction of the momenta of
%% the final state objects since the observable requires the evaluation
%% of the invariant mass squared $\sttj$.  
The unfolding performed in
\Ref{Aad:2015waa} is based on a modeling of various effects including
estimates on the related uncertainties.  Furthermore, it has been
shown in \Ref{Aad:2015waa} that the unfolding is, within the
uncertainties, independent of the top-quark mass used in the unfolding
procedure. Instead of performing the unfolding, a future mass
determination could also employ more realistic theoretical predictions
reducing or even avoiding the unfolding. As long as the uncertainties
of the unfolding procedure are reliably estimated, we do not expect a
major difference in the results, although the related uncertainties in
the mass determination may be reduced using improved theoretical
predictions. In particular, relying on a complete theoretical
description of top-quark production and decay, including off-shell
effects, using for example the recently published results on $e^+\nu_e\mu^- \bar
\nu_\mu b \bar b j +X$ production
\cite{Bevilacqua:2015qha,Bevilacqua:2016jfk} may  
reduce the theoretical uncertainties. As far as the combination
with the parton shower is concerned, the recent developments on parton
showers at NLO accuracy, taking into account intermediate resonances
\cite{Jezo:2015aia,Jezo:2016ujg}, may also help to further improve the
theoretical predictions. 

\section{Proof of concept: determination of the top-quark running mass}
\label{sec:massdetermination}
In this section we use the theoretical results presented in the
previous section to extract the top-quark \MSbar mass from the
experimental results published in \Ref{Aad:2015waa}.

As shown in \Refs{Alioli:2013mxa,Aad:2015waa}, the most sensitive bin
is the interval $0.675 < \rhos < 1$. Very close to the kinematic
threshold ($\rhos \sim 1$), fixed-order calculations as presented in
the previous section are most likely not sufficient, since bound-state
effects and soft gluon emission may become important
\cite{Hagiwara:2008df,Kiyo:2008bv}. However, as has been shown in
\Refs{Hagiwara:2008df,Kiyo:2008bv} the impact on the total cross
section is small. As far as the $\ttbar$ invariant mass distribution
is concerned a significant distortion due to (would-be) bound-state
effects occurs only below or closely above the nominal threshold.  In
\Ref{Aad:2015waa} the impact of the threshold region on the mass
extraction has been investigated using different upper boundaries. As
no significant variation of the extracted mass value was found, the
range for the highest $\rhos$-bin has been extended to 1 in
\Ref{Aad:2015waa}.  We assume that the same holds true for the \MSbar
mass and use the same bin boundaries for the bin close to the
threshold. This assumption is well justified given the current uncertainties.

\begin{figure}
\includegraphics[width=\columnwidth]{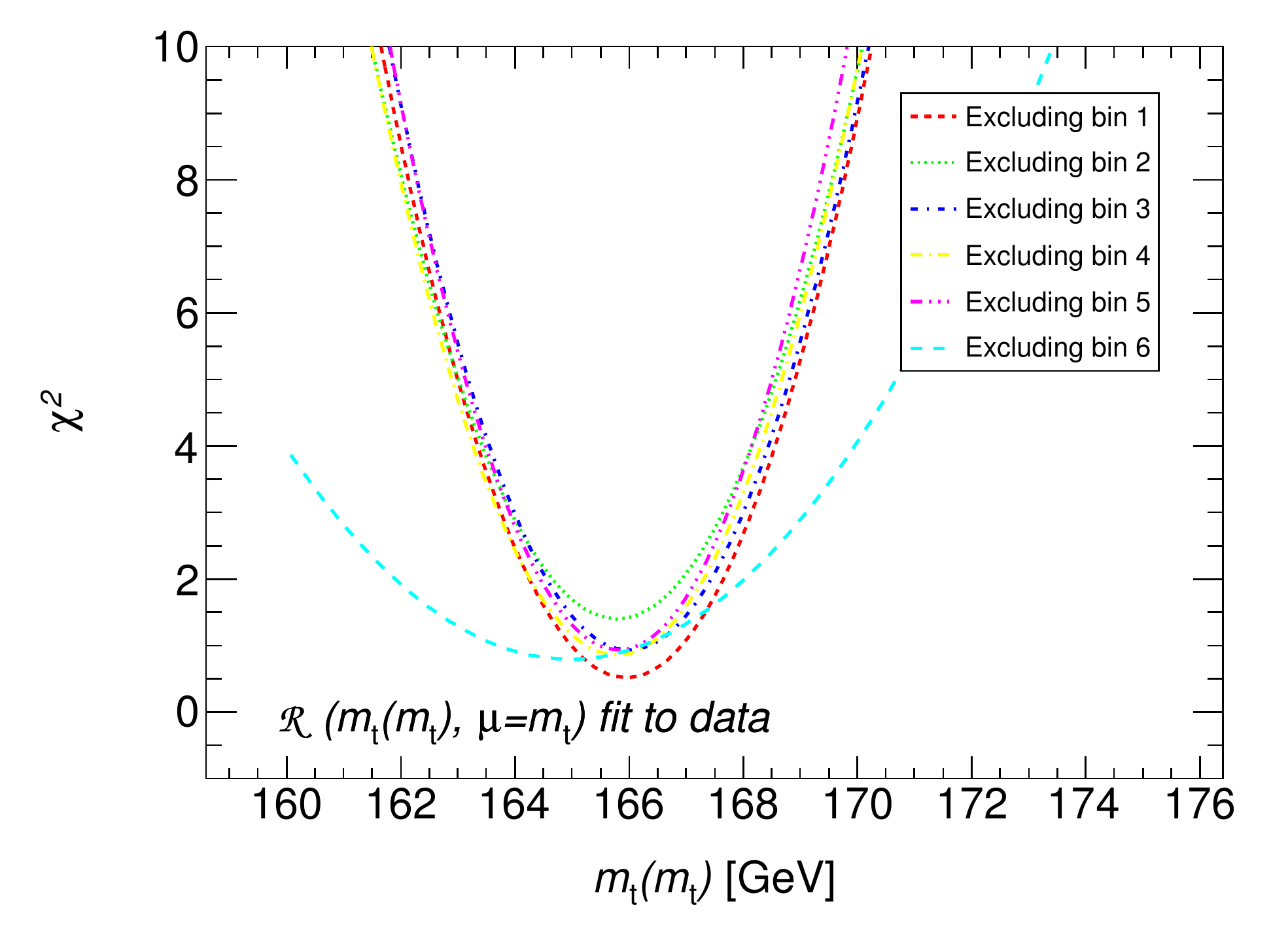}
\caption{The $\chi^2$ as a function of \mMS for 
        different choices of the excluded bin. The bin boundaries for
        the six bins are given in Table \ref{tab:n3_scale}. \label{fig:chi}}
\end{figure}

In complete analogy with what has been done in \Ref{Aad:2015waa} we use
a least-square fit and define the $\chi^2$ by
\begin{equation}
  \chi^{2}=\sum \limits_{ij} \left[\n3^{\corr,i}-\n3^{i}(\mMS) \right]
  V^{-1}_{ij}\left[\n3^{\corr,j}-\n3^{j}(\mMS)\right],
  \label{eq:chi}
\end{equation}
where $\n3^{\corr,i}$ is the measured value in the $i$-th
bin - unfolded to the parton-level - and $\n3^{i}(\mMS)$ denotes the
theoretical prediction as a function of the running mass. Since the
numerical evaluation of $\n3^{i}(\mMS)$ is very time consuming we
calculate $\n3^{i}(\mMS)$ for $\mMS$ in the range 160--170 GeV in
steps of one GeV and use a linear interpolation in between.  The
matrix $V^{-1}$ is the inverse of the statistical covariance matrix of
the unfolded \n3-distribution as given in \Ref{Aad:2015waa} (Figure 8
provided as auxiliary material to \Ref{Aad:2015waa}).  The top-quark
mass value is determined through a minimization of the $\chi^2$. We
use only five of the six available bins because the number of degrees
of freedom is reduced by one through the normalization of the $\n3$
distribution. Figure \ref{fig:chi} shows the results for different choices
of the excluded bin.  As long as the bin closest to the threshold is
kept, very similar results are obtained. Excluding the highest bin,
however, leads to an important change of the observed $\chi^2$. This
is a direct consequence of the high sensitivity of the cross section
in this bin to the top-quark mass together with the small experimental
uncertainty of the measured value for this bin. The exclusion of the
highest bin leads to a shift of about 1 GeV as far as the minimum of
the $\chi^2$ is concerned.  In addition, the minimum becomes
significantly broader, reflecting the loss in sensitivity.
For the extraction of the top-quark mass we follow the approach used
in \Ref{Aad:2015waa} and  exclude the lowest bin which corresponds to
the most energetic events and has only a very weak sensitivity to the
top-quark mass. The statistical uncertainty is taken as the mass
shift that increases $\chi^{2}$ by one unit with respect to the 
minimum ($\Delta \chi^2 =$ +1).

As the published results of \Ref{Aad:2015waa} do not provide a full
covariance matrix, including the breakdown of the systematic
uncertainties, we use a modified version of the $\chi^2$ function where
no covariance matrix is assumed to evaluate the systematic
uncertainties:
\begin{equation}
  \chi^{2}=\sum \limits_{i} 
  \frac{\left[\n3^{\corr,i}-\n3^{i}(\mMS)\right]^2}{\n3^i(\mMS)}.
  \label{eq:chi2}
\end{equation}

The fit is repeated 5000 times with pseudo data varying the value of
$\n3^{\corr,i}$ assuming a gaussian distribution with a mean value
(standard deviation) for $\n3^{\corr,i}$ taken from the first (last)
column of Tab.~4 of \Ref{Aad:2015waa}.  As a consistency check the
average of the different mass values obtained is calculated.  Its
value agrees within 0.1 GeV with the value obtained using \Eq{eq:chi}.
The systematic uncertainties are estimated as the standard deviation
of the individual fit results.

A full implementation of the systematic effects is thus missing in our
study, however, we expect that the results given here represent a
reasonable approximation.  
The theoretical uncertainties are evaluated by repeating the fit with
different choices for the PDF and renormalization and factorization
scales. The PDF uncertainty is taken as the maximum difference between
the nominal result and the mass obtained with MSTW2008nlo90cl and
NNPDF2.3 NLO. It has only a minor effect on the mass: $\Delta \mMS
=0.15~\GeV$. The scale variation is determined as the asymmetric
difference between the mass extracted with the central scale choice,
$\mu=\mMS$, and with the scales $\mur$, $\muf$ set to
$\mur=\muf=2\mMS$ and $\mur=\muf=\mMS/2$. The two
theory uncertainties are added in quadrature. For the determination of
the aforementioned uncertainties, we used the theoretical \n3 distribution
calculated with $\mu=\mMS=166~\GeV$ as pseudo data instead of 
$\n3^{\corr,i}$ as determined in \Ref{Aad:2015waa} to avoid 
that statistical fluctuations of the experimental data 
introduces a bias in the evaluation of the theoretical
uncertainties.

As a final result we obtain for the running
top-quark mass:
\begin{eqnarray*}
  \mMS &=&165.9 \pm 1.4 ~({\rm stat.}) \pm 1.3 ~({\rm syst.}) ^{+
  1.5}_{-0.6} ~({\rm theory})~{\rm GeV},
\end{eqnarray*}
where the central value and the statistical uncertainty are evaluated
using \Eq{eq:chi} while the systematic and theoretical
uncertainties are determined using \Eq{eq:chi2} as described above.
We have not included the uncertainty due to the parametric uncertainty
of the QCD coupling constant because it has been shown in
\Ref{Aad:2015waa} that the impact is negligible. 
Comparing with \Ref{Aad:2015waa}, we observe a slight decrease 
of the experimental uncertainties. 
Combining the individual uncertainties in quadrature, we find
\begin{eqnarray*}
  \mMS &=&165.9 ^{+ 2.4}_{-2.0} ~({\rm total})~{\rm GeV}.
\end{eqnarray*}
Comparing with the result for the top-quark pole mass as given in
\Ref{Aad:2015waa}
\begin{eqnarray*}
  \mtPole = 173.7\pm 1.5 \mbox{ (stat.)} \pm 1.4 \mbox{ (syst.)}
  ^{+1.0}_{-0.5}\mbox{ (theory) GeV \cite{Aad:2015waa}}, 
\end{eqnarray*}
we find that the theoretical uncertainties are larger
in the extraction of \mMS. In both cases the theoretical uncertainty
is largely dominated by the scale variation while the PDF effects give
only a small contribution. Inspecting the individual bins it turns out
that the bin closest to the threshold is responsible for the larger
scale dependence when using the \MSbar mass. 
Most likely this is due to the appearance of the
derivative in \Eq{eq:SchemeConversion}, which can lead to a sizaable
contribution in the threshold region where $d\sigma/dm$ is large.
Given the high sensitivity with respect to the top-quark mass this
leads to an important contribution to the scale variation of the
extracted mass value. We also note that with respect to the behavior
of the perturbative expansion the two schemes, pole mass and running
mass, are on an equal footing: As can be seen from \Eq{eq:M1} no large
logarithms are involved in the relation between \mtPole and \mMS. The
slight improvement of the scale dependence in the inclusive $t\bar t$ cross
sections is to a large extent a kinematic effect, since the
numerically smaller mass value in the running mass scheme increases the
Born approximation and thus reduces the size of the corrections, leading to
a somewhat smaller scale variation. As explained above, distributions
may show a different behavior. 
 
As a consistency check of the determination of the running mass one can
convert the extracted mass value into the
pole mass scheme and vice versa. Since in both cases the quality of
the perturbative expansion is comparable we do not expect any major
differences. The result of this exercise is shown in
Table \ref{table_masses}.
\begin{table}[h]
  \begin{center}
    \begin{tabular}{c|ccc}
      \hline
      {\hfill Theory} & {\hfill \mMS (GeV)} & {} & 
      {\hfill \mtPole (GeV)}   \\
      \hline
      \n3(\mMS)  & \hfill $165.9 ^{+ 2.4}_{-2.0}$ &
      \hfill $\rightarrow$      &\hfill $173.5 ^{+ 2.5}_{-2.1}$  \\
      \n3(\mtPole)  & \hfill $165.8 ^{+ 2.2}_{-2.0}$ &
      \hfill $\leftarrow$      &\hfill $173.7 ^{+ 2.3}_{-2.1}$  \\\hline
    \end{tabular}
  \end{center}
  \caption{Top-quark mass extracted in the pole mass
    scheme and converted to the running mass scheme and vice versa 
    \label{table_masses}
    }
\end{table} 
Indeed the results are in perfect agreement with each other. Note that
we included only the first non-trivial term in $\alpha_s$ in the
conversion. This is a consistent approximation within the NLO accuracy
presented here. Furthermore, no uncertainty due to the mass conversion itself
is considered. 

\section{Conclusion}
\label{sec:conclusion}
In this article the NLO QCD corrections for top-quark pair production
in association with an additional jet have been calculated  defining the
top-quark mass in the \MSbar scheme instead of the top-quark pole mass
commonly used. As a proof of concept, we have shown that it is possible
to determine the top-quark running mass from the measurement of the 
differential cross section of $\ttbaronejet$ production. Using the data
from \Ref{Aad:2015waa} we find
\begin{eqnarray*}
  \mMS &=&165.9 \pm 1.4 ~({\rm stat.}) \pm 1.3 ~({\rm syst.}) ^{+
  1.5}_{-0.6} ~({\rm theory})~{\rm GeV},
\end{eqnarray*}
which is - when translated to the pole mass scheme - consistent with
the measurement of the pole mass. Since we extract the running
top-quark mass at the scale of the top-quark mass itself, the
conversion between the two mass schemes does not involve large
logarithms. As a consequence, a major improvement concerning the
convergence of the perturbative expansion is not expected. Using \mMS
is appropriate for the measurements reported in \Ref{Aad:2015waa}
which is based on 7 TeV data collected in run I. Due to the limited
statistics, the moderate collider energy, and the large sensitivity of
the bin closest to the threshold, high energetic events ($\rho_s <
0.4$) have only little impact on the analysis. Setting the
renormalization scale equal to the top-quark mass and using \mMS
should thus provide reliable results. This is supported by the good
agreement between the mass measurement using the pole mass and the
mass measurement using the running mass. The larger theory uncertainty
when using the running mass is due to the high mass sensitivity of the
differential cross section close to the threshold and its larger scale
dependence. As one can read off from Table \ref{tab:n3_scale} the scale
variation of the sixth bin is roughly twice as large as the scale
variation in bins three, four, and five. (The scale variation of bin
one and two is similar or even larger than that of bin six, however,
due to the small number of events (about 10\% of the total number of
events) and the reduced sensitivity, these two bins do not
significantly affect the
outcome of the analysis.)  

%%%% With larger statistics in run II one may
%%%% restrict the analysis to one bin in the threshold region with
%%%% optimized bin boundaries.  Lowering the upper bin boundary may result
%%%% in a smaller scale dependence at the price of slightly reducing the
%%%% mass sensitivity. This could be partially compensated by reducing the
%%%% lower bin boundary which would increase the number of events and
%%%% should thus decrease the statistical uncertainty.
%% Note that because of the high mass sensitivity this would need
%% to be balanced by more precise measurements at smaller $\rhos$. 
%% In general, independent whether the pole mass or the \MSbar mass is
%% measured, the increased energy and the higher luminosity in run II may
%% help in reducing the theoretical uncertainties which are dominated by
%% the scale uncertainty, since the impact of the bin closest to the threshold
%% which is to a large extent responsible for the scale uncertainty may
%%  be reduced using more high energetic events.
%%% The increased energy and the higher luminosity in run II may help in
%%% reducing the theoretical uncertainties which are dominated by the
%%% scale uncertainty since the impact of events with $\rhos$ close to the
%%% threshold which is to a large extent responsible for the scale
%%% uncertainty may be reduced using more high energetic events. 
We note that the measurement of the running mass gives an interesting
and complementary description to that obtained with the pole mass.
With more high energetic events available in run II it could
also be interesting to determine $\mMSmu$ where the renormalization
scale is set to some large value reflecting the typical momentum
transfer in the considered $\rhos$ bin (in the case that several bins are used). This would give the
opportunity to `measure' the running of the top-quark mass.
Furthermore, at large scales the perturbative expansion using the
running top-quark mass may be improved since the conversion introduces
now potentially large logarithms which may cancel corresponding terms
explicit in the calculation. However, given the significantly smaller
sensitivity for low $\rhos$ large statistics and a very good control
of the systematic uncertainties would be required.

%
% -------------------------------------------------------------------------

\section*{Acknowledgments}

We would like to thank T. Carli and F. Deliot for their careful
 reading of the manuscript and their valuable comments.
This work is partially supported by the  German Federal
Ministry for Education and Research (Grant 05H15KHCAA), the 
Spanish Ministry of Economy, Industry and Competitiveness
(MINEICO/FEDER-UE, FPA2015-65652-C4-3-R) and by the Marie Curie
Initial Training Network of the European Union
HiggsTools-ITN, No. 316704, FP7-PEOPLE-2012-ITN.

\bibliographystyle{utphys}
\bibliography{refs}

\providecommand{\href}[2]{#2}\begingroup\raggedright\begin{thebibliography}{10}

\bibitem{Abe:1995hr}
{\bfseries CDF} Collaboration, F.~Abe {\em et~al.}, ``{Observation of top quark
  production in $\bar{p}p$ collisions},'' {\em Phys.Rev.Lett.} {\bfseries 74}
  (1995) 2626--2631,
\href{http://arxiv.org/abs/hep-ex/9503002}{{\ttfamily arXiv:hep-ex/9503002
  [hep-ex]}}.
%%CITATION = HEP-EX/9503002;%%.

\bibitem{Abachi:1995iq}
{\bfseries D0} Collaboration, S.~Abachi {\em et~al.}, ``{Observation of the top
  quark},'' {\em Phys.Rev.Lett.} {\bfseries 74} (1995) 2632--2637,
\href{http://arxiv.org/abs/hep-ex/9503003}{{\ttfamily arXiv:hep-ex/9503003
  [hep-ex]}}.
%%CITATION = HEP-EX/9503003;%%.

\bibitem{ATLAS:2014wva}
{\bfseries ATLAS, CDF, CMS, D0} Collaboration, ``{First combination of Tevatron
  and LHC measurements of the top-quark mass},''
\href{http://arxiv.org/abs/1403.4427}{{\ttfamily arXiv:1403.4427 [hep-ex]}}.
%%CITATION = ARXIV:1403.4427;%%".

\bibitem{Degrassi:2012ry}
G.~Degrassi, S.~Di~Vita, J.~Elias-Miro, J.~R. Espinosa, G.~F. Giudice, {\em
  et~al.}, ``{Higgs mass and vacuum stability in the Standard Model at NNLO},''
  \href{http://dx.doi.org/10.1007/JHEP08(2012)098}{{\em JHEP} {\bfseries 1208}
  (2012) 098},
\href{http://arxiv.org/abs/1205.6497}{{\ttfamily arXiv:1205.6497 [hep-ph]}}.
%%CITATION = ARXIV:1205.6497;%%.

\bibitem{Alekhin:2012py}
S.~Alekhin, A.~Djouadi, and S.~Moch, ``{The top quark and Higgs boson masses
  and the stability of the electroweak vacuum},''
  \href{http://dx.doi.org/10.1016/j.physletb.2012.08.024}{{\em Phys.Lett.}
  {\bfseries B716} (2012) 214--219},
\href{http://arxiv.org/abs/1207.0980}{{\ttfamily arXiv:1207.0980 [hep-ph]}}.
%%CITATION = ARXIV:1207.0980;%%.

\bibitem{Cortiana:2015rca}
G.~Cortiana, ``{Top-quark mass measurements: review and perspectives},''
  \href{http://dx.doi.org/10.1016/j.revip.2016.04.001}{{\em Rev. Phys.}
  {\bfseries 1} (2016) 60--76},
\href{http://arxiv.org/abs/1510.04483}{{\ttfamily arXiv:1510.04483 [hep-ex]}}.
%%CITATION = ARXIV:1510.04483;%%".

\bibitem{Vos:2016tof}
{\bfseries ATLAS, CMS} Collaboration, M.~Vos, ``{Top-quark mass measurements at
  the LHC: alternative methods},'' {\em PoS} {\bfseries TOP2015} (2016) 035,
\href{http://arxiv.org/abs/1602.00428}{{\ttfamily arXiv:1602.00428 [hep-ex]}}.
%%CITATION = ARXIV:1602.00428;%%".

\bibitem{Hoang:2000yr}
A.~H. Hoang {\em et~al.}, ``{Top - anti-top pair production close to threshold:
  Synopsis of recent NNLO results},''
  \href{http://dx.doi.org/10.1007/s1010500c0003}{{\em Eur. Phys. J.direct}
  {\bfseries C3} (2000) 1--22},
\href{http://arxiv.org/abs/hep-ph/0001286}{{\ttfamily arXiv:hep-ph/0001286
  [hep-ph]}}.
%%CITATION = HEP-PH/0001286;%%".

\bibitem{Bigi:1994em}
I.~I.~Y. Bigi, M.~A. Shifman, N.~G. Uraltsev, and A.~I. Vainshtein, ``{The Pole
  mass of the heavy quark. Perturbation theory and beyond},''
  \href{http://dx.doi.org/10.1103/PhysRevD.50.2234}{{\em Phys. Rev.} {\bfseries
  D50} (1994) 2234--2246},
\href{http://arxiv.org/abs/hep-ph/9402360}{{\ttfamily arXiv:hep-ph/9402360}}.
%%CITATION = HEP-PH/9402360;%%".

\bibitem{Beneke:1994sw}
M.~Beneke and V.~M. Braun, ``{Heavy quark effective theory beyond perturbation
  theory: Renormalons, the pole mass and the residual mass term},''
  \href{http://dx.doi.org/10.1016/0550-3213(94)90314-X}{{\em Nucl. Phys.}
  {\bfseries B426} (1994) 301--343},
\href{http://arxiv.org/abs/hep-ph/9402364}{{\ttfamily arXiv:hep-ph/9402364}}.
%%CITATION = HEP-PH/9402364;%%".

\bibitem{Beneke:2016cbu}
M.~Beneke, P.~Marquard, P.~Nason, and M.~Steinhauser, ``{On the ultimate
  uncertainty of the top quark pole mass},''
\href{http://arxiv.org/abs/1605.03609}{{\ttfamily arXiv:1605.03609 [hep-ph]}}.
%%CITATION = ARXIV:1605.03609;%%".

\bibitem{Langenfeld:2009wd}
U.~Langenfeld, S.~Moch, and P.~Uwer, ``{Measuring the running top-quark
  mass},'' \href{http://dx.doi.org/10.1103/PhysRevD.80.054009}{{\em Phys.Rev.}
  {\bfseries D80} (2009) 054009},
\href{http://arxiv.org/abs/0906.5273}{{\ttfamily arXiv:0906.5273 [hep-ph]}}.
%%CITATION = ARXIV:0906.5273;%%.

\bibitem{Alekhin:2016jjz}
S.~Alekhin, S.~Moch, and S.~Thier, ``{Determination of the top-quark mass from
  hadro-production of single top-quarks},''
\href{http://arxiv.org/abs/1608.05212}{{\ttfamily arXiv:1608.05212 [hep-ph]}}.
%%CITATION = ARXIV:1608.05212;%%".

\bibitem{Alioli:2013mxa}
S.~Alioli, P.~Fernandez, J.~Fuster, A.~Irles, S.-O. Moch, P.~Uwer, and M.~Vos,
  ``{A new observable to measure the top-quark mass at hadron colliders},''
  \href{http://dx.doi.org/10.1140/epjc/s10052-013-2438-2}{{\em Eur. Phys. J.}
  {\bfseries C73} (2013) 2438},
  \href{http://arxiv.org/abs/1303.6415}{{\ttfamily arXiv:1303.6415 [hep-ph]}}.

\bibitem{Aad:2015waa}
{\bfseries ATLAS} Collaboration, ``{Determination of the top-quark pole mass
  using $ t\overline{t} $ + 1-jet events collected with the ATLAS experiment in
  7 TeV pp collisions},'' \href{http://dx.doi.org/10.1007/JHEP10(2015)121}{{\em
  JHEP} {\bfseries 10} (2015) 121},
\href{http://arxiv.org/abs/1507.01769}{{\ttfamily arXiv:1507.01769 [hep-ex]}}.
%%CITATION = ARXIV:1507.01769;%%".

\bibitem{CMS:2016khu}
{\bfseries CMS} Collaboration, ``{Determination of the normalised invariant
  mass distribution of $\mathrm{t\bar{t}}+$jet and extraction of the top quark
  mass},''
\href{http://arxiv.org/abs/CMS-PAS-TOP-13-006}{{\ttfamily CMS-PAS-TOP-13-006}}.
%%CITATION = CMS-PAS-TOP-13-006;%%".

\bibitem{Dittmaier:2007wz}
S.~Dittmaier, P.~Uwer, and S.~Weinzierl, ``{NLO QCD corrections to t anti-t +
  jet production at hadron colliders},''
  \href{http://dx.doi.org/10.1103/PhysRevLett.98.262002}{{\em Phys.Rev.Lett.}
  {\bfseries 98} (2007) 262002},
\href{http://arxiv.org/abs/hep-ph/0703120}{{\ttfamily arXiv:hep-ph/0703120
  [HEP-PH]}}.
%%CITATION = HEP-PH/0703120;%%.

\bibitem{Dittmaier:2008uj}
S.~Dittmaier, P.~Uwer, and S.~Weinzierl, ``{Hadronic top-quark pair production
  in association with a hard jet at next-to-leading order QCD: Phenomenological
  studies for the Tevatron and the LHC},''
  \href{http://dx.doi.org/10.1140/epjc/s10052-008-0816-y}{{\em Eur.Phys.J.}
  {\bfseries C59} (2009) 625--646},
\href{http://arxiv.org/abs/0810.0452}{{\ttfamily arXiv:0810.0452 [hep-ph]}}.
%%CITATION = ARXIV:0810.0452;%%.

\bibitem{Melnikov:2009dn}
K.~Melnikov and M.~Schulze, ``{NLO QCD corrections to top quark pair production
  and decay at hadron colliders},''
  \href{http://dx.doi.org/10.1088/1126-6708/2009/08/049}{{\em JHEP} {\bfseries
  08} (2009) 049},
\href{http://arxiv.org/abs/0907.3090}{{\ttfamily arXiv:0907.3090 [hep-ph]}}.
%%CITATION = ARXIV:0907.3090;%%".

\bibitem{Melnikov:2010iu}
K.~Melnikov and M.~Schulze, ``{NLO QCD corrections to top quark pair production
  in association with one hard jet at hadron colliders},''
  \href{http://dx.doi.org/10.1016/j.nuclphysb.2010.07.003}{{\em Nucl. Phys.}
  {\bfseries B840} (2010) 129--159},
\href{http://arxiv.org/abs/1004.3284}{{\ttfamily arXiv:1004.3284 [hep-ph]}}.
%%CITATION = ARXIV:1004.3284;%%".

\bibitem{Cacciari:2008gp}
M.~Cacciari, G.~P. Salam, and G.~Soyez, ``{The Anti-k(t) jet clustering
  algorithm},'' \href{http://dx.doi.org/10.1088/1126-6708/2008/04/063}{{\em
  JHEP} {\bfseries 0804} (2008) 063},
\href{http://arxiv.org/abs/0802.1189}{{\ttfamily arXiv:0802.1189 [hep-ph]}}.
%%CITATION = ARXIV:0802.1189;%%.

\bibitem{Cacciari:2011ma}
M.~Cacciari, G.~P. Salam, and G.~Soyez, ``{FastJet user manual},''
  \href{http://dx.doi.org/10.1140/epjc/s10052-012-1896-2}{{\em Eur.Phys.J.}
  {\bfseries C72} (2012) 1896},
\href{http://arxiv.org/abs/1111.6097}{{\ttfamily arXiv:1111.6097 [hep-ph]}}.
%%CITATION = ARXIV:1111.6097;%%.

\bibitem{Nadolsky:2008zw}
P.~M. Nadolsky {\em et~al.}, ``{Implications of CTEQ global analysis for
  collider observables},'' {\em Phys. Rev.} {\bfseries D78} (2008) 013004,
  \href{http://arxiv.org/abs/arXiv:0802.0007}{{\ttfamily arXiv:0802.0007}}.

\bibitem{Martin:2009iq}
A.~Martin, W.~Stirling, R.~Thorne, and G.~Watt, ``{Parton distributions for the
  LHC},'' \href{http://dx.doi.org/10.1140/epjc/s10052-009-1072-5}{{\em
  Eur.Phys.J.} {\bfseries C63} (2009) 189--285},
\href{http://arxiv.org/abs/0901.0002}{{\ttfamily arXiv:0901.0002 [hep-ph]}}.
%%CITATION = ARXIV:0901.0002;%%.

\bibitem{Ball:2012cx}
R.~D. Ball {\em et~al.}, ``{Parton distributions with LHC data},''
  \href{http://dx.doi.org/10.1016/j.nuclphysb.2012.10.003}{{\em Nucl. Phys.}
  {\bfseries B867} (2013) 244--289},
\href{http://arxiv.org/abs/1207.1303}{{\ttfamily arXiv:1207.1303 [hep-ph]}}.
%%CITATION = ARXIV:1207.1303;%%".

\bibitem{Bevilacqua:2015qha}
G.~Bevilacqua, H.~B. Hartanto, M.~Kraus, and M.~Worek, ``{Top Quark Pair
  Production in Association with a Jet with Next-to-Leading-Order QCD Off-Shell
  Effects at the Large Hadron Collider},''
  \href{http://dx.doi.org/10.1103/PhysRevLett.116.052003}{{\em Phys. Rev.
  Lett.} {\bfseries 116} no.~5, (2016) 052003},
\href{http://arxiv.org/abs/1509.09242}{{\ttfamily arXiv:1509.09242 [hep-ph]}}.
%%CITATION = ARXIV:1509.09242;%%".

\bibitem{Bevilacqua:2016jfk}
G.~Bevilacqua, H.~B. Hartanto, M.~Kraus, and M.~Worek, ``{Off-shell Top Quarks
  with One Jet at the LHC: A comprehensive analysis at NLO QCD},''
  \href{http://dx.doi.org/10.1007/JHEP11(2016)098}{{\em JHEP} {\bfseries 11}
  (2016) 098},
\href{http://arxiv.org/abs/1609.01659}{{\ttfamily arXiv:1609.01659 [hep-ph]}}.
%%CITATION = ARXIV:1609.01659;%%".

\bibitem{Jezo:2015aia}
T.~Ježo and P.~Nason, ``{On the Treatment of Resonances in Next-to-Leading
  Order Calculations Matched to a Parton Shower},''
  \href{http://dx.doi.org/10.1007/JHEP12(2015)065}{{\em JHEP} {\bfseries 12}
  (2015) 065},
\href{http://arxiv.org/abs/1509.09071}{{\ttfamily arXiv:1509.09071 [hep-ph]}}.
%%CITATION = ARXIV:1509.09071;%%".

\bibitem{Jezo:2016ujg}
T.~Ježo, J.~M. Lindert, P.~Nason, C.~Oleari, and S.~Pozzorini, ``{An NLO+PS
  generator for ${{t \bar{t}}}$ and ${{W t}}$ production and decay including
  non-resonant and interference effects},''
  \href{http://dx.doi.org/10.1140/epjc/s10052-016-4538-2}{{\em Eur. Phys. J.}
  {\bfseries C76} no.~12, (2016) 691},
\href{http://arxiv.org/abs/1607.04538}{{\ttfamily arXiv:1607.04538 [hep-ph]}}.
%%CITATION = ARXIV:1607.04538;%%".

\bibitem{Hagiwara:2008df}
K.~Hagiwara, Y.~Sumino, and H.~Yokoya, ``{Bound-state Effects on Top Quark
  Production at Hadron Colliders},''
  \href{http://dx.doi.org/10.1016/j.physletb.2008.07.006}{{\em Phys. Lett.}
  {\bfseries B666} (2008) 71--76},
\href{http://arxiv.org/abs/0804.1014}{{\ttfamily arXiv:0804.1014 [hep-ph]}}.
%%CITATION = ARXIV:0804.1014;%%".

\bibitem{Kiyo:2008bv}
Y.~Kiyo, J.~H. Kuhn, S.~Moch, M.~Steinhauser, and P.~Uwer, ``{Top-quark pair
  production near threshold at LHC},''
  \href{http://dx.doi.org/10.1140/epjc/s10052-009-0892-7}{{\em Eur. Phys. J.}
  {\bfseries C60} (2009) 375--386},
\href{http://arxiv.org/abs/0812.0919}{{\ttfamily arXiv:0812.0919 [hep-ph]}}.
%%CITATION = ARXIV:0812.0919;%%".

\end{thebibliography}\endgroup

\section*{Appendix}

See Tables \ref{tab:n3_pdf_nnpdf} and \ref{tab:n3_pdf_mstw}.
%%Table with theoretical R-values -- MSTW2008
\begin{table}[!htbp]
\begin{center}\renewcommand{\arraystretch}{1.2}
\begin{tabular}{l|l|l|l} 
\hline
& \multicolumn{3}{c}{$\n3(\mMS)$ }\\
\hline
bin, range in $\rhos$ & $160$ GeV & $161$ GeV &  $162$ GeV   \\
\hline                                                            
1, 0-0.25   & ${0.115}$ &    ${0.117}$    &    ${0.118}$\\     
2, 0.25-0.325  & ${1.057}$ &    ${1.056}$    &    ${1.060}$\\     
3, 0.325-0.425 & ${1.918}$ &    ${1.940}$    &    ${1.968}$\\     
4, 0.425-0.525 & ${2.484}$ &    ${2.520}$    &    ${2.525}$\\     
5, 0.525-0.675  & ${2.163}$ &    ${2.151}$    &    ${2.157}$\\     
6, 0.675-1.0    & ${0.392}$ &    ${0.378}$    &    ${0.364}$\\
\hline
\end{tabular}
\begin{tabular}{c|l|l|l|l} 
\multicolumn{4}{c}{}\\
\multicolumn{4}{c}{}\\
\hline
 bin& $163$ GeV &  $164$ GeV &  $165$ GeV &  $166$ GeV \\
\hline                                                            
1&${0.120}$   &    ${0.122}$    &    ${0.124}$    &    ${0.127}$ \\    
2&${1.088}$   &    ${1.096}$    &    ${1.115}$    &    ${1.131}$ \\    
3&${1.971}$   &    ${2.018}$    &    ${2.036}$    &    ${2.048}$       \\      
4&${2.546}$   &    ${2.567}$    &    ${2.592}$    &    ${2.602}$       \\      
5&${2.153}$   &    ${2.131}$    &    ${2.132}$    &    ${2.127}$       \\      
6&${0.350}$   &    ${0.336}$    &    ${0.317}$    &    ${0.306}$       \\
\hline
\end{tabular}
\begin{tabular}{c|l|l|l|l} 
\multicolumn{4}{c}{}\\
\multicolumn{4}{c}{}\\
\hline
bin& $167$ GeV &  $168$ GeV &  $169$ GeV &  $170$ GeV  \\
\hline                                                            
1&${0.128}$   &    ${0.130}$    &    ${0.132}$    &    ${0.136}$ \\    
2&${1.143}$   &    ${1.155}$    &    ${1.179}$    &    ${1.194}$ \\    
3&${2.082}$   &    ${2.104}$    &    ${2.127}$    &    ${2.154}$ \\    
4&${2.619}$   &    ${2.645}$    &    ${2.640}$    &    ${2.644}$ \\    
5&${2.119}$   &    ${2.107}$    &    ${2.103}$    &    ${2.097}$ \\    
6&${0.290}$   &    ${0.276}$    &    ${0.266}$    &    ${0.253}$ \\
\hline
\end{tabular}
\end{center}
\caption{Same as Table \ref{tab:n3_scale}  but for 
  the NNPDF2.3 PDF set. Only the central scale $\mu=\mMS$ is shown.
  \label{tab:n3_pdf_nnpdf}}
\end{table}

%%Table with theoretical R-values -- MSTW2008
\begin{table}[!htbp]
\begin{center}\renewcommand{\arraystretch}{1.2}
\begin{tabular}{l|l|l|l} 
\hline
 & \multicolumn{3}{c}{$\n3(\mMS)$ }\\
\hline
bin, range in $\rhos$   & $160$ GeV & $161$ GeV &  $162$ GeV   \\
\hline                                                            
1, 0-0.25   & ${0.119}$ &    ${0.120}$    &    ${0.124}$\\     
2, 0.25-0.325  & ${1.045}$ &    ${1.065}$    &    ${1.072}$\\     
3, 0.325-0.425 & ${1.919}$ &    ${1.926}$    &    ${1.968}$\\     
4, 0.425-0.525 & ${2.460}$ &    ${2.490}$    &    ${2.521}$\\     
5, 0.525-0.675  & ${2.163}$ &    ${2.156}$    &    ${2.144}$\\     
6, 0.675-1.0    & ${0.398}$ &    ${0.385}$    &    ${0.364}$\\
\hline
\end{tabular}
\begin{tabular}{c|l|l|l|l} 
\multicolumn{4}{c}{}\\
\multicolumn{4}{c}{}\\
\hline
Bin& $163$ GeV &  $164$ GeV &  $165$ GeV &  $166$ GeV \\
\hline                                                            
1&${0.124}$   &    ${0.128}$    &    ${0.128}$    &    ${0.130}$ \\    
2&${1.094}$   &    ${1.097}$    &    ${1.119}$    &    ${1.144}$ \\    
3&${1.994}$   &    ${2.003}$    &    ${2.027}$    &    ${2.047}$       \\      
4&${2.527}$   &    ${2.545}$    &    ${2.567}$    &    ${2.580}$       \\      
5&${2.140}$   &    ${2.154}$    &    ${2.131}$    &    ${2.130}$       \\      
6&${0.351}$   &    ${0.332}$    &    ${0.323}$    &    ${0.306}$       \\
\hline
\end{tabular}
\begin{tabular}{c|l|l|l|l} 
\multicolumn{4}{c}{}\\
\multicolumn{4}{c}{}\\
\hline
Bin& $167$ GeV &  $168$ GeV &  $169$ GeV &  $170$ GeV  \\
\hline                                                            
1&${0.136}$   &    ${0.136}$    &    ${0.136}$    &    ${0.141}$ \\    
2&${1.157}$   &    ${1.167} $    &    ${1.182}$    &    ${1.208}$ \\    
3&${2.070}$   &    ${2.091} $    &    ${2.115}$    &    ${2.133}$ \\    
4&${2.588}$   &    ${2.612} $    &    ${2.629}$    &    ${2.649}$ \\    
5&${2.124}$   &    ${2.115} $    &    ${2.110}$    &    ${2.089}$ \\    
6&${0.292}$   &    ${0.280} $    &    ${0.266}$    &    ${0.255}$ \\
\hline
\end{tabular}
\end{center}
\caption{Same as Table \ref{tab:n3_scale}  but for 
  the MSTW2008nlo PDF set. Only the central scale $\mu=\mMS$ is shown
  \label{tab:n3_pdf_mstw}}
\end{table}

\end{document}